\documentclass[twocolumn, switch]{article}
\usepackage{preprint}

\usepackage[utf8]{inputenc} 
\usepackage[T1]{fontenc}    
\usepackage{hyperref}
\usepackage{comment}

\usepackage{url}            
\usepackage{booktabs}       
\usepackage{bm}
\usepackage{amsfonts}       
\usepackage{amsmath}
\usepackage{amssymb}
\usepackage{nicefrac}       
\usepackage{microtype}
\usepackage{cleveref}
\usepackage{xcolor}
\usepackage{algpseudocode}
\usepackage{algorithm}
\usepackage{color,soul}
\usepackage{caption}
\usepackage{lipsum}
\usepackage{dsfont}
\usepackage{breqn}

\usepackage{amsmath,amssymb}
\usepackage{fancyhdr}       
\usepackage{graphicx}       
\graphicspath{{media/}}     

\makeatletter
\newcommand{\printfnsymbol}[1]{%
  \textsuperscript{\@fnsymbol{#1}}%
}

\newcommand*\bigcdot{\mathpalette\bigcdot@{.5}}
\newcommand*\bigcdot@[2]{\mathbin{\vcenter{\hbox{\scalebox{#2}{$\m@th#1\bullet$}}}}}

\makeatother

\usepackage{graphicx} 

\title{VoroTO: Multiscale Topology Optimization of Voronoi Structures using Surrogate Neural Networks}

\author{
\and 
Rahul Kumar Padhy \\
Department of Mechanical Engineering \\
University of Wisconsin-Madison \\
Madison, WI, USA \\
\texttt{rkpadhy@wisc.edu} \\
\and 
Krishnan Suresh \\
Department of Mechanical Engineering \\
University of Wisconsin-Madison \\
Madison, WI, USA \\
\texttt{ksuresh@wisc.edu} \\
\and
Aaditya Chandrasekhar \\
Department of Mechanical Engineering \\
University of Wisconsin-Madison \\
Madison, WI, USA \\
\texttt{cs.aaditya@gmail.com} \\
}

\begin{document}

\twocolumn[\begin{@twocolumnfalse}
\maketitle

\begin{abstract}

Cellular structures found in nature exhibit remarkable properties such as high strength, high energy absorption, excellent thermal/acoustic insulation, and fluid transfusion. Many of these structures are Voronoi-like; therefore researchers have proposed Voronoi multi-scale designs for a wide variety of engineering applications. However, designing such structures can be computationally prohibitive due to the multi-scale nature of the underlying analysis and optimization.
In this work, we propose the use of a neural network (NN) to carry out efficient topology optimization (TO) of multi-scale Voronoi structures. The NN is first trained using Voronoi parameters (cell site locations, thickness, orientation, and anisotropy) to predict the homogenized constitutive properties. This network is then integrated into a conventional TO framework to minimize structural compliance subject to a volume constraint. Special considerations are given for ensuring positive definiteness of the constitutive matrix and promoting macroscale connectivity. Several numerical examples are provided to showcase the proposed method.

\end{abstract}


 \vspace{1cm}
 
\end{@twocolumnfalse}]

\begin{figure*}
 	\begin{center}
		\includegraphics[scale=0.7,trim={0 0 0 0},clip]{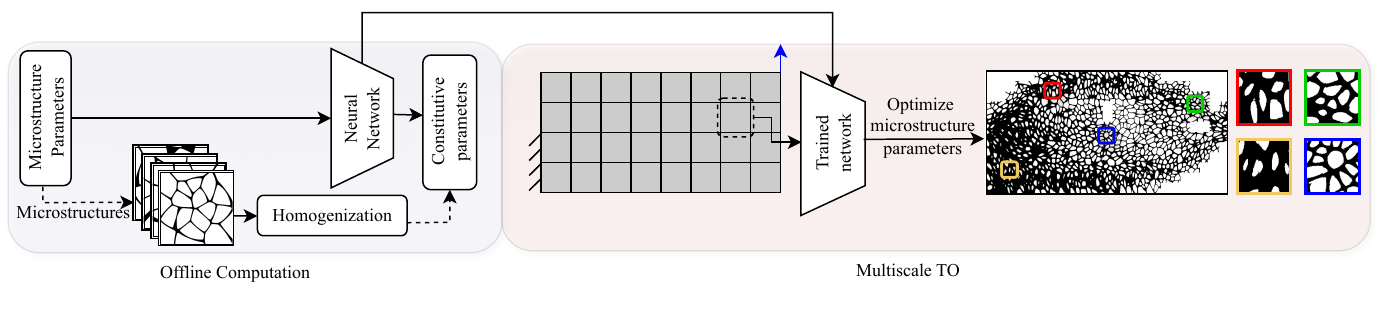}
 		\caption{Graphical abstract: Offline computation: Given a dataset containing Voronoi microstructure parameters and homogenized constitutive properties, a neural network is trained offline. Multiscale TO: The trained network is used as a surrogate during topology optimization to derive optimized Voronoi structures. }
        \label{fig:voroto_graphical_abstract}
	\end{center}
 \end{figure*}

 \vspace{1cm}
 
\section{Introduction}
\label{sec:intro}

Topology optimization is a powerful engineering tool employed to optimize the distribution of material within a given design domain to achieve optimal performance under specified constraints \cite{bendsoe2013topology, sigmund2013topology}. Traditionally, single-scale topology optimization focuses on determining the optimal layout of material within a structure at a single scale, aiming primarily at maximizing stiffness \cite{bendsoe2013topology} for a given set of loading conditions.
However, engineering applications often demand more than just structural efficiency \cite{gibson2010review}. Features such as structural robustness \cite{knoll2009design}, high strength \cite{hu2017design}, superior energy absorption \cite{ebrahimi2022plane}, fluid circulation \cite{cardoso2013advances}, thermal and acoustic insulation \cite{ashby1997cellular} capabilities are crucial. These attributes not only enhance the overall performance of structures but also facilitate multifunctionality, making them indispensable across various industries, ranging from aerospace \cite{klippstein2018additive, wu2022controllable}, biomedical \cite{hollister2005porous, seol2013new}, and more \cite{chen2023functionally}.

\begin{figure}
 	\begin{center}
		\includegraphics[scale=1.1,trim={0 0 0 0},clip]{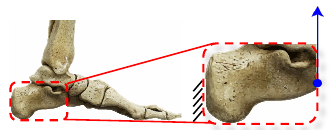}
 		\caption{Heel bone and loading conditions.}
        \label{fig:bone}
	\end{center}
 \end{figure}

\begin{figure}[H]
\centering
\hspace{0.8cm}\includegraphics[width=.22\textwidth]{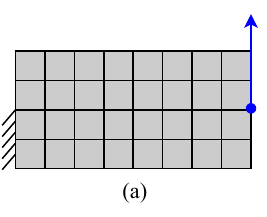}\quad
\includegraphics[width=.25\textwidth]{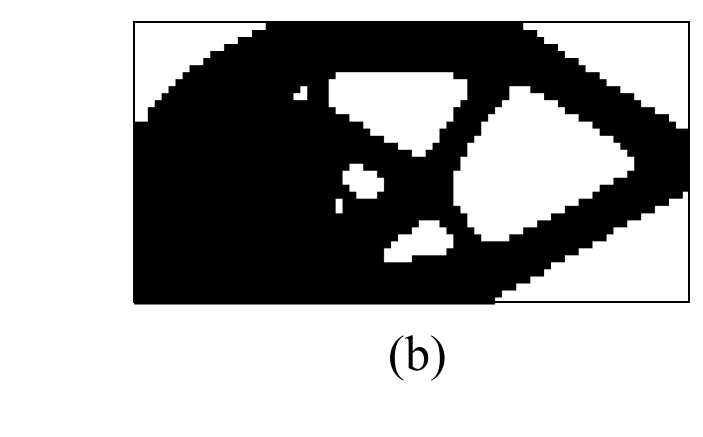}\quad
\includegraphics[width=.25\textwidth]{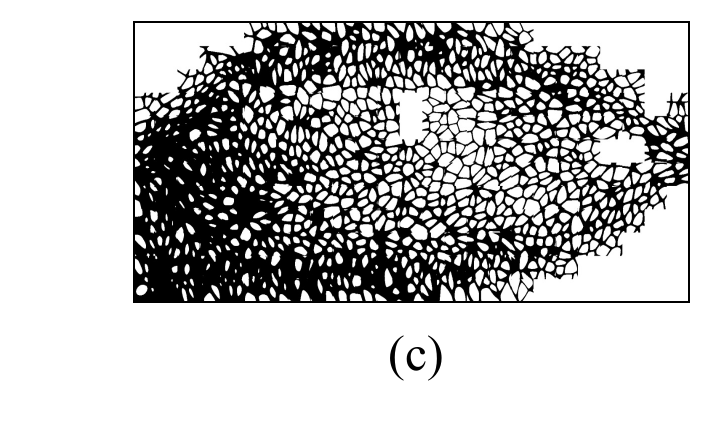}

    \caption{(a) Topology optimization problem.  (b) Single scale design.  (c) A multiscale porous design. }
\label{fig:comparison_single_voro}
\end{figure}

 Many natural structures, such as bones \cite{gautam2021nondestructive}, wood \cite{ufodike2021additively}, and insect wings \cite{jongerius2010structural} exhibit remarkable combinations of strength, energy absorption, fluid circulation, and insulating properties. These features are often achieved through the intricate arrangement of porous structures, which possess desirable properties such as stochasticity \cite{zaharin2018effect}, anisotropy \cite{jacobs1997adaptive, park2018design}, and connectivity \cite{bala2016pore}. For instance, consider the calcaneus (heel bone) and its associated loading conditions during walking, as depicted in \cref{fig:bone}. This scenario can be effectively modeled as a simple cantilever beam as in \cref{fig:comparison_single_voro}{a} \cite{park2018design}. The growth of bone entails a continuous self-optimization process, where its internal structure adapts to maximize mechanical efficiency under varying external conditions. This behavior can be modeled through topology optimization, by formulating the problem of maximizing stiffness \cite{goda2019topology, park2018design}. 

The heel bone topology optimization problem is idealized \cref{fig:comparison_single_voro}{a}. An optimal design that maximizes stiffness using single-scale topology optimization is presented in \cref{fig:comparison_single_voro}{b}. Although this design is optimal for stiffness, it lacks the porosity essential for fluid movement. In bones, pores play a vital role in blood and interstitial fluid movement, enabling the exchange of oxygen and nutrients between tissues \cite{cardoso2013advances}. A hypothetical porous structure that sacrifices stiffness for porosity \cite{nguyen2022lightweight, syam2018design, valdevit2011protocols} is depicted in  \cref{fig:comparison_single_voro}{c}.

Systematic computation of such structures poses significant computational challenges due to their multiscale nature \cite{lei2014multi, martinez2016procedural}. Theoretically, one can use high-resolution single-scale topology optimization to arrive at such structures, but this is not viable in practice \cite{martinez2016procedural}. Here, we present a neural network (NN) based computationally efficient multiscale topology optimization approach for computing such structures. Furthermore, we use Voronoi microstructures to represent porosity as they offer high design freedom \cite{wang2018design}, exhibit anisotropy, and often resemble the porous structures found in bone \cite{audibert2018bio, ying2018anisotropic}. A parametric representation of the microstructures is used to train an NN, facilitating the mapping of microstructure parameters to homogenized properties. Special attention is given to the parametric representation to promote macroscale connectivity, addressing a limitation \cite{wu2021topology} not inherently present in unit cell-based multiscale topology optimization approaches \cite{padhy2023fluto, padhy2023tomas}.

\subsection{Topology optimization of porous structures}
\label{sec:lit_pores}

Topology optimization of porous structures has gained considerable attention, with two primary approaches \cite{wu2021topology}: full-scale and multi-scale. The full-scale approach employs geometrical constraints such as pattern repetition \cite{wu2015system, wu2019design} and local volume constraints \cite{guest2009imposing, dou2020projection, das2020multi} to generate optimized porous structures. Recent enhancements include incorporating length scale constraint \cite{zhao2021design} and parametrized porous structures \cite{feng2023cellular}. Although these approaches can generate optimized porous structures, the incorporation of additional geometrical constraints often results in reduced performance when compared to traditional topology optimization methods \cite{lu2023designing}. Furthermore, these approaches can be computationally expensive because of the need for high-resolution topology optimization of fine porous structures \cite{martinez2016procedural}. 

Multi-scale approaches separate the optimization problem into macro and micro scales, enabling customization of local material properties and macroscale structural behavior \cite{wu2021topology, zhao2023concurrent}. It has yielded materials with exceptional properties, such as micro-modules with high Young's modulus \cite{bendsoe1999material}. Advanced techniques such as clustering \cite{yan2020clustering} and level-set-based \cite{li2016integrated} optimization are used to enhance structural properties and to obtain designs with smooth boundaries respectively. Fixed types of parametric cells were used to design porous structures such as planar rod networks \cite{schumacher2018mechanical}, or TPMS (triply periodic minimal surfaces) based \cite{tozoni2020low,yan2019strong, hu2020efficient}. These approaches reduce computational costs in generating foams with valid geometry but have a limited design space \cite{li2023explicit}. For a comprehensive review of these approaches, please see \cite{wu2021topology}.

\subsection{Voronoi based porous structure optimization}
\label{sec:lit_voronoi}
Traditionally, regular shapes such as diamond \cite{jette2018femoral} and honeycomb \cite{ghorbani20203d} have been used to represent porous structures. However, these regular structures are susceptible to stress shielding and often exhibit limited permeability for fluid movement \cite{chao2023evaluation}. Consequently, there is growing research interest in irregular porous structures, specifically, Voronoi tessellation \cite{lei2020parametric, wang2018design}, to model the complex shapes of bone structures.  Voronoi tessellations offer high design freedom \cite{wang2018design}, exhibit anisotropy, and often resemble the porous structures found in nature \cite{audibert2018bio, ying2018anisotropic}.

Various stochastic methods based on Voronoi tessellations have been proposed \cite{lu2014build, kou2010simple, martinez2016procedural}, but these lead to extensive data occupancy and are computationally intensive \cite{wang2023stochastic}. An alternative method involves utilizing seed positions and beam radii as design parameters, enabling them to conform to freeform shapes while maintaining geometric connectivity. However, this does not facilitate the generation of anisotropic designs \cite{li2023explicit}. Another technique employs an iterative process of removing and adding site points to achieve a desired stress distribution on the porous structure, but it lacks the capability for optimizing a Voronoi structure through continuous sensitivity analysis \cite{cucinotta2019stress}. A multi-scale approach was introduced for designing Voronoi-graded cellular structures for heat transfer problems. This method adapts offline computation to reduce computational expenses \cite{chen2023multiscale}. Nevertheless, the resulting design lacks anisotropy and has a constant Voronoi strut thickness across the entire design domain.

\subsection{Contributions}
\label{sec:intro_contribution}

This paper presents a novel framework for generating optimized Voronoi multiscale designs with spatially varying thickness and anisotropy.  In contrast to full-scale approach for Voronoi design generation \cite{feng2023cellular}, we leverage a multiscale design paradigm, utilizing an offline-trained NN that maps Voronoi microstructure parameters to homogenized constitutive microstructure properties. Key contributions of this work include:  

\begin{itemize}

    \item \textbf{Design space:} We consider anisotropic Voronoi cells of varying thickness in our design space, enhancing performance \cite{ying2018anisotropic}.

    \item \textbf{Connectivity:} In the process of designing Voronoi microstructures, neighboring cell sites are taken into account to promote connectivity.
    
    \item \textbf{Physically valid:} We ensure that the stiffness matrix predicted by the network is positive definite, i.e., physically valid.

    \item \textbf{End-to-end differentiability:} Our methodology is end-to-end differentiable, enabling automated sensitivity analysis for gradient-based optimization.

\end{itemize}

\section{Proposed Method}
\label{sec:method}

\subsection{Voronoi Diagram and Structure}
\label{sec:method_voronoi}
\begin{figure*}
 	\begin{center}
		\includegraphics[scale=0.8,trim={0 0 0 0},clip]{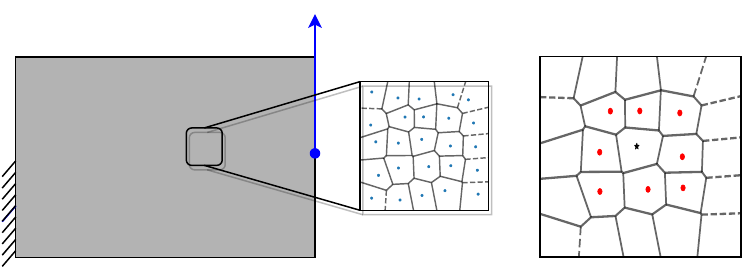}
 		\caption{ Voronoi diagram defined by a set of sites (points). The shape of cell ($\bigstar$) is influenced by its immediate neighbors (\textcolor{red}{$\bigcdot$}).}
        \label{fig:macroscale_voronoi}
	\end{center}
 \end{figure*}

Our objective here is to design porous structures using Voronoi diagrams.  A Voronoi diagram is a geometric structure that partitions space into regions or cells based on the proximity to a specified set of points called sites \cite{lee1981generalization, choset2000sensor, hoff1999fast}; see \cref{fig:macroscale_voronoi}. The shape of the Voronoi cell associated with a site is only influenced by that site and nearby sites.  Voronoi diagrams are valuable for representing and analyzing complex spatial relationships, making them essential in computational geometry \cite{arseneva2019randomized, sainlot2017restricting}, modeling of porous structures \cite{lei2020parametric}, and design problems \cite{feng2023cellular}. 

Using the Voronoi \emph{diagram} as a basis, we will now define a Voronoi \emph{structure}. Let the domain contain $\tilde{S}$  sites (points), denoted by $ (\bar x_s, \bar y_s); s = 1,2, ... , \tilde{S} $; this results in $\tilde{S}$ cells. We now associate 3 parameters with each cell: (1) $\beta_s$ to create a thickness to the cells, (2)  $\alpha_s$ to control the degree of anisotropy, and (3) $\theta_s$ to control the orientation. These parameters are essential to obtain an optimal distribution of porous material. The resulting Voronoi structure is defined using a density function as follows. Given any point $ (x,y) $  we define an anisotropic \cite{labelle2003anisotropic, van2021anisotropic} and oriented distance \cite{gusrialdi2008voronoi} of $ (x,y) $ to any cell site $s$ as:

\begin{equation}
    d_s(x, y) = \sqrt{\alpha_s (\delta_s^x) ^2 + (\delta_s^y)^2/\alpha_s }
    \label{eq:dist_site_point}
\end{equation}
where,
\begin{equation}
    \begin{pmatrix}
    \delta_s^x\\
    \delta_s^y
    \end{pmatrix} = 
    \begin{bmatrix}
        \cos \theta_s & -\sin \theta_s \\
        \sin \theta_s & \cos \theta_s
    \end{bmatrix}
    \begin{pmatrix}
        x - \bar x_s \\
        y - \bar y_s
    \end{pmatrix}
\label{eq:voronoi_rotated_dist}
\end{equation}

Then the density field associated with a site $s$ is defined via a soft-max function \cite{feng2023cellular}:

 \begin {equation}
    \hat{\rho}_s(x, y) = \bigg(\frac{e^{ -k d_s(x,y)}}{\sum\limits_{s=1}^{S}e^{-k d_s(x,y)}}\bigg)^{\beta_s}
    \label{eq:elem_site_encoding}
\end{equation}

where the parameter $k$ controls the sharpness of the soft-max function \cite{feng2023cellular}. Observe that $\hat{\rho}_s(\bar x_s, \bar y_s) = 1$ when the point ($x$, $y$) is closest to site $s$, and approximately 0 far away from it.  Finally, we can compute the total density field at ($x$, $y$) as:

\begin{equation}
    \rho(x, y ) = 1 -  \sum\limits_{s=1}^{\tilde{S}} \hat{\rho}_s(x,y)
    \label{eq:final_dens}
\end{equation}

\begin{figure}
 	\begin{center}
		\includegraphics[scale=0.9,trim={0 0 0 0},clip]{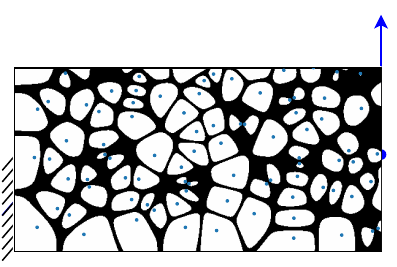}
 		\caption{A typical density field defining a Voronoi structure. } 
        \label{fig:fullscale_voronoi}
	\end{center}
 \end{figure}

 A typical density field is illustrated in \cref{fig:fullscale_voronoi}. While the sites control the topology of the Voronoi structure, the parameters $\beta_s$, $\alpha_s$, and $\theta_s$ control its shape; additional examples are provided later.

\subsection{Single Scale Voronoi Structure Optimization}
\label{sec:single_scale}
In principle, one can now optimize the location of the $\tilde{S}$ cell sites $(\bar{x}_s, \bar{y}_s)$, the thicknesses $\beta_{s}$, anisotropies $\alpha_{s}$ and orientations $\theta_{s}$ to design optimal Voronoi structures. For example, one may pose a compliance minimization problem, subject to a volume constraint as:

\begin{subequations}
	\label{eq:optimization_base_Eqn_single_scale}
	\begin{align}
		& \underset{\bar{\bm{x}}, \bar{\bm{y}}, \bm{\beta}, \bm{\theta}, \bm{\alpha}} {\text{minimize}}  & J = \bm{f}^T \bm{u} \label{eq:opt_sing_scale_objective} \\
		& \text{subject to}
		& \bm{K}\bm{u} = \bm{f}\label{eq:opt_sing_scale_govEq}\\
		& \text{and} &  V \leq V^{max}  
            \label{eq:opt_single_scale_base_volCons} \\
		& &  x_{min} \leq \bar{x}_s \leq x_{max} \; , \; \forall s \\ 
            & &  y_{min} \leq \bar{y}_s \leq y_{max} \; , \; \forall s \\
            & &  \beta_{min} \leq \beta_s  \leq \beta_{max}  \; , \; \forall s \label{eq:opt_single_scale_beta_bounds} \\
            & &  \alpha_{min} \leq \alpha_s  \leq \alpha_{max} \; , \; \forall s \label{eq:opt_single_scale_alpha_bounds} \\
            & &  \theta_{min} \leq \theta_s  \leq \theta_{max}  \; , \; \forall s \label{eq:opt_single_scale_theta_bounds}
	\end{align}
\end{subequations}
where $\bm{K}$ is global stiffness matrix, $\bm{f}$  is the applied load. For example, the authors of \cite{feng2023cellular} illustrate optimizing Voronoi structures by varying cell site locations, thickness, and anisotropy to minimize compliance. However, this is computationally prohibitive for designing fine-scale porous structures since it will require a large number of cell sites $\tilde{S}$, but, more importantly, it will entail a dense finite element mesh to capture the thin features. In other words, a single-scale optimization is not a viable strategy for achieving porous structures. Instead, we will pursue a multi-scale framework, discussed in the next section.

\subsection{Proposed Method: Overview}
\label{sec:prob_form}
\begin{figure*}
 	\begin{center}
		\includegraphics[scale=0.8,trim={0 0 0 0},clip]{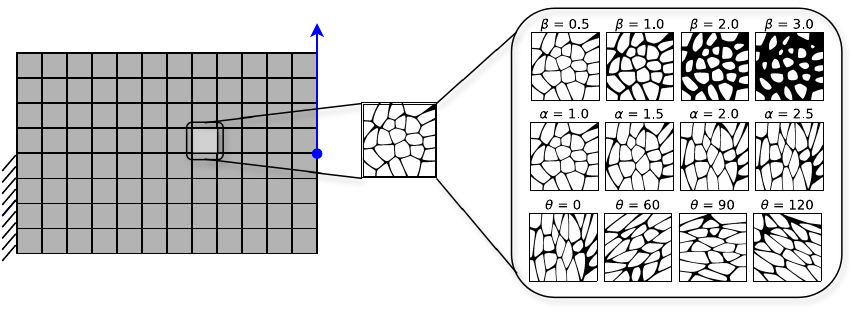}
 		\caption{Given a discretized domain, we wish to populate each element with Voronoi cells. All cells in a particular microstructure have the same thickness ($\beta$), and are anisotropic ($\alpha$) with an orientation ($\theta$).}
        \label{fig:voro_cell_variations}
	\end{center}
 \end{figure*}

To achieve a multi-scale design, we discretize the domain into a finite number of \emph {macro} elements, where each element $e$ contains a small number of $S$ cell sites,  $(\bar{x}^{(e)}_{s}, \bar{y}^{(e)}_{s})$; $s = 1, 2, ..., S$; $S = 4$ in this paper. Furthermore, each element will be associated with 3 additional parameters: the thickness parameter ${\beta}^{(e)}$, the anisotropy parameter ${\alpha}^{(e)}$, and the orientation parameter ${\theta}^{(e)}$. Observe that the parameters associated with an element are common to all sites within the element. By optimizing all parameters over all elements (as discussed in the remainder of the paper), one can design a multi-scale porous structure; this is schematically illustrated in \cref{fig:voro_cell_variations}. However, direct optimization will still be computationally prohibitive due to the need for high resolution finite element analysis.

To reduce the computational cost we propose a two-stage strategy (see the graphical abstract \cref {fig:voroto_graphical_abstract}):

\begin{enumerate}
    \item \textbf{Offline Computation}: This involves generating numerous representative Voronoi microstructures, computing their homogenized properties and training of NNs (see \cref{sec:offline}).
    \item \textbf{Multi-Scale Optimization}: Exploiting the trained NNs to carry out efficient multi-scale Voronoi structure generation (see \cref{sec:multiscale_optimization}). 
\end{enumerate}


\subsection{Offline Computation}
\label{sec:offline}
We now describe various steps in the offline computation.
\subsubsection{Voronoi Microstructure Generation}
\label{sec:method_dataset}

\begin{figure*}
 	\begin{center}
		\includegraphics[scale=0.45,trim={0 0 0 0},clip]{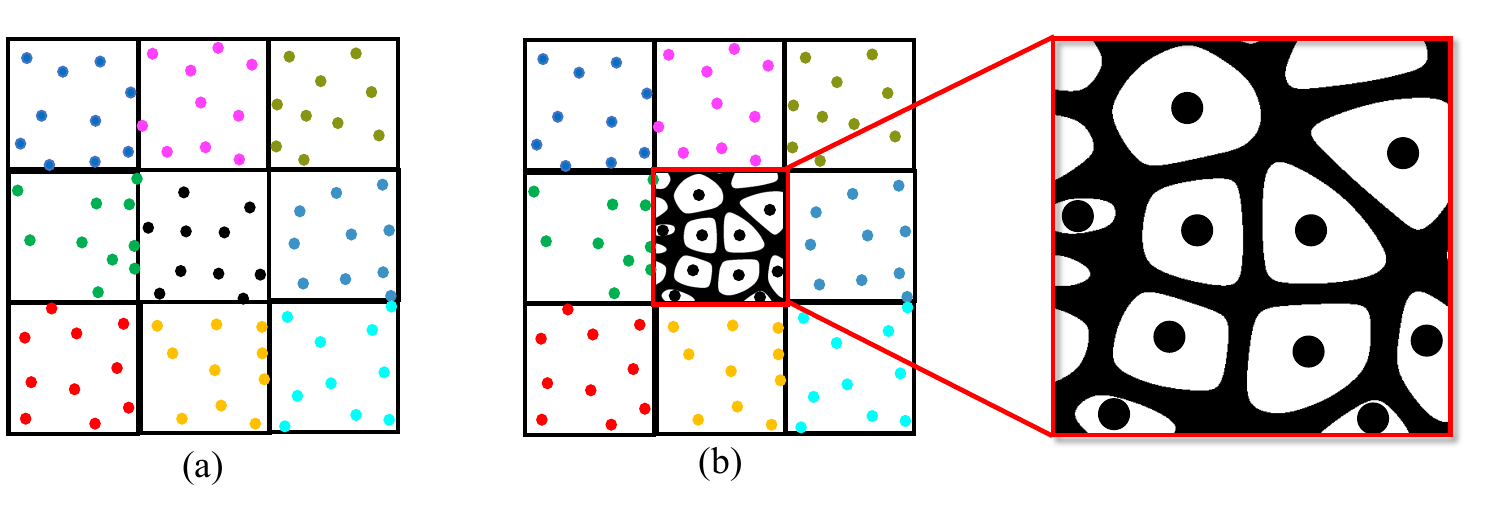}
 		\caption{Data generation involves (a) distributing Voronoi cell site locations for the central microstructure element (1) and
its neighbors (8), (b) generating Voronoi microstructures based on cell locations, thickness, degree of anisotropy and rotation values for the central microstructure element  (c) extracting the central microstructure design.}
        \label{fig:data_gen}
	\end{center}
 \end{figure*}

The first step is to create a large, representative set of Voronoi microstructures associated with the macro elements. A naive approach is to generate $S$ random sites (points) within a macro element and generate the corresponding Voronoi microstructure. However, this would not represent a typical microstructure since points in the neighboring macro elements also influence the Voronoi microstructure. For example, the naive approach would fail to generate solid material on the element's boundary. We therefore consider a macro element and its 8 neighboring macro elements, and generate random cell sites in all 9 elements; see \cref{fig:data_gen}a. Then, a Voronoi microstructure is generated in the central element using all the cell sites, and a random set of parameters \cite{oliphant2006guide} $\beta$, $\alpha$ and $\theta$, that are generated uniformly over a pre-defined range, as described in numerical experiments; see \cref{fig:data_gen}b for a typical Voronoi microstructure.

\begin{figure}
 	\begin{center}
		\includegraphics[scale=0.15,trim={0 0 0 0},clip]{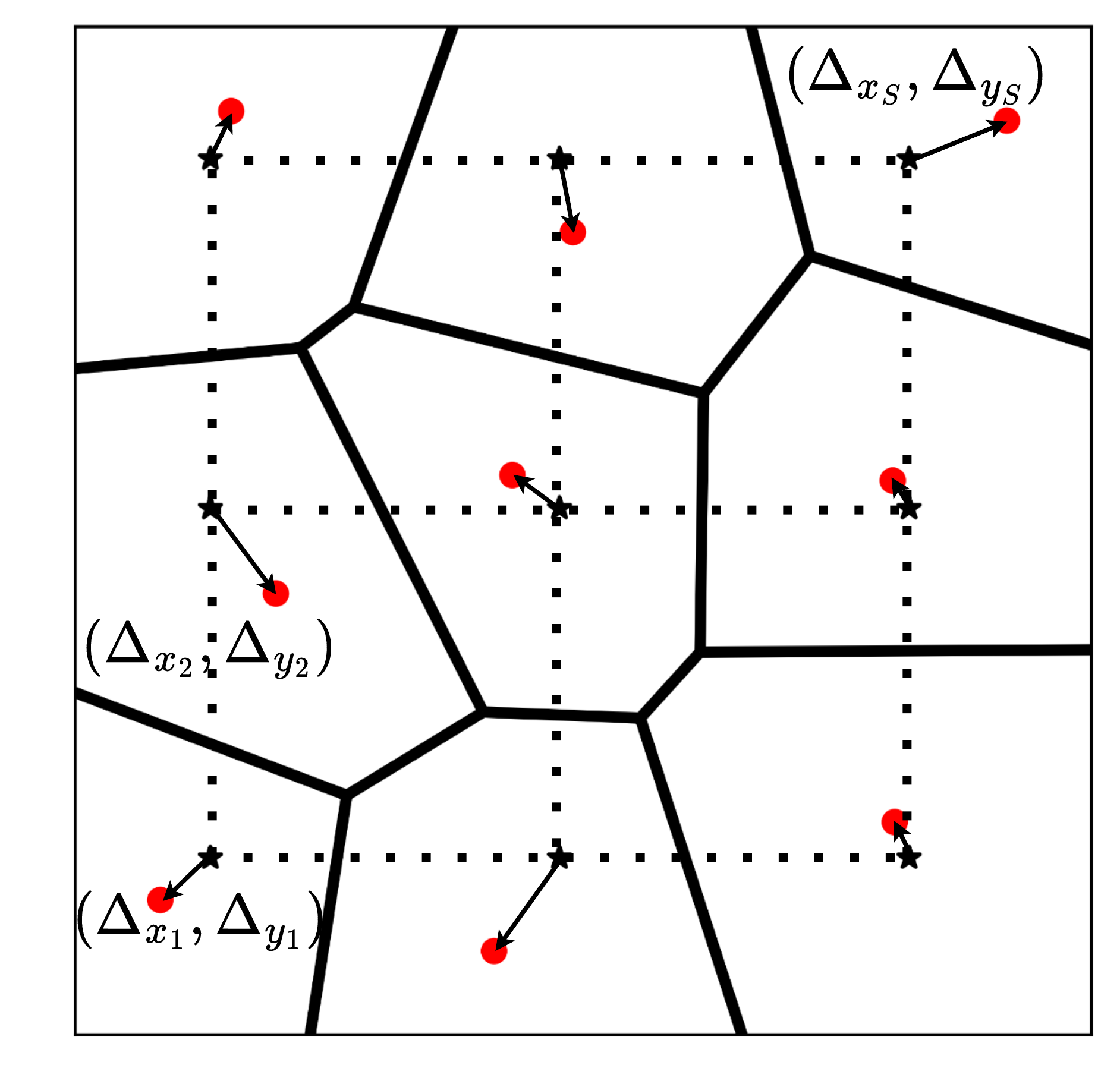}
 		\caption{The cell sites $(\mathcolor{red}\bigcdot)$ within a microstructure are obtained as a perturbation to a grid of points $(\boldsymbol{\bigstar})$.}
        \label{fig:cell_generation_perturbation}
	\end{center}
 \end{figure}

Further, to ensure that the points are sufficiently separated from one another, we model the points as a perturbation $(\Delta_x, \Delta_y)$ from grid points; see \cref{fig:cell_generation_perturbation}. 

\subsubsection{Homogenization}
\label{sec:homogeniz}
\begin{figure*}
 	\begin{center}
		\includegraphics[scale=0.45,trim={0 0 0 0},clip]{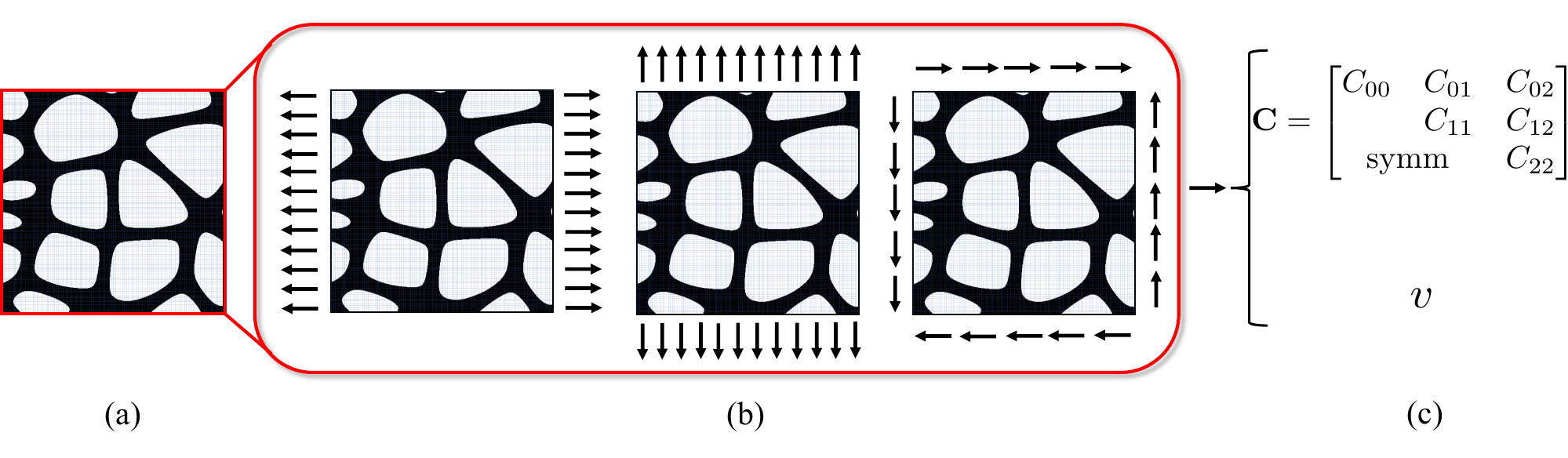}
 		\caption{Numerical homogenization:(a) Discretization of the density field onto a mesh, (d) employing numerical homogenization to determine the (c) constitutive properties of the microstructure.}
        \label{fig:homogen}
	\end{center}
 \end{figure*}

 Once these Voronoi microstructures are generated, the macro element is discretized into a grid of $120 \times 120$ \emph {micro} elements; see \cref{fig:homogen}a. The mesh is then used for numerical homogenization to compute the elasticity matrix $\mathbf{C}$ and the volume fraction $v$; see \cref{fig:homogen}b and \cref{fig:homogen}c.  Without a loss of generality, we assume a Young's Modulus $E=1$ and a Poisson's ratio $\nu = 0.3$; see \cite{andreassen2014howToHomogenize} for details on numerical homogenization.

\subsubsection{Cholesky Decomposition}
\label{sec:chol_decom}

In the subsequent sections, we employ a NN to establish a mapping between the Voronoi parameters, and the corresponding homogenized elasticity matrix components (and volume fraction). However, a direct mapping might lead to the prediction of stiffness matrices that are not positive definite \cite{jekel2022neural}. To avoid this, we first carry out a Cholesky decomposition of $\mathbf C$:

\begin{equation}
    \mathbf{C} =\mathbf{L} \mathbf{L}^{T}
    \label{eq:chol_decom}
\end{equation}

where,
\[
\mathbf{L}  = 
\begin{bmatrix}
  L_{00} & 0 & 0 \\
  L_{10} & L_{11} & 0 \\
  L_{20} & L_{21} & L_{22} \\
\end{bmatrix}
\]
The components of $\mathbf{L}$, i.e.,  $\{L_{00}, L_{10} ..., L_{22}\}$ are then used for training the NN, as discussed next.

\subsubsection{Neural Network Training} 
\label{sec:sur_nn}

Following the data generation and Cholesky decomposition, we use NNs \cite{hornik1989multilayer} to establish a mapping between the Voronoi parameters and computed data. NNs have been utilized for various material modeling applications, including modeling the plastic response of metal \cite{furukawa2004accurate}, to construct a relation between macroscopic stress and crack opening responses \cite{unger2009neural},  homogenization of composite structures \cite{lefik2009artificial, man2011neural}, constitutive modeling of elastomeric foams \cite{liang2008neural}, and nonlinear response of carbon nanotubes \cite{papadopoulos2018neural}. In addition, NNs have been utilized in the context of multi-scale topology optimization \cite{white2019multiscale}. 

For the design of microstructures using NNs pixel-based representations have been employed \cite{wang2020data, wang2020deep, kudyshev2020machine}. This method assumes designs can be constructed using solid or void elements in space. While this allows extensive design flexibility, it has drawbacks \cite{lee2024data}. 
There is a considerable computational burden for both design evaluation and machine learning due to limitations in scalability with respect to resolution \cite{lee2024data}. Second, efficient exploration of the design space requires the utilization of constraints to achieve desired design attributes \cite{ma2019probabilistic}. 
Many studies \cite{kollmann2020deep, zhao2021machine, qiu2019deep, malkiel2018plasmonic} employing NNs trained on pixel-based representations to predict microstructure properties typically optimize the networks to minimize the difference between the predicted design and a ground truth solution using loss functions such as mean squared error (MSE), mean absolute error (MAE), or binary cross-entropy. However, this approach may result in inaccurate predictions as structurally similar designs in a pixelated format can exhibit significantly different properties \cite{lee2024data, woldseth2022use}.

\begin{figure}[H]
 	\begin{center}
		\includegraphics[scale=1.,trim={0 0 0 0},clip]{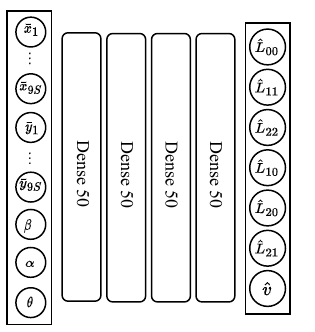}
 		\caption{Proposed surrogate neural network architecture. }
        \label{fig:nn_architecture}
	\end{center}
 \end{figure}

Here, NNs are employed to predict the homogenized elastic response (and volume fraction) as a continuous and differentiable mapping of microstructure parameters, facilitating gradient-based optimization \cite{chandrasekhar2021tounn}. The proposed NN architecture consists of the following components (see \cref{fig:nn_architecture}):

\begin{enumerate}
    \item The input consists of the $(x,y)$ location of the sites (points) from all 9 macro elements (the elements are ordered bottom to top, then left to right, and  the parameters $\beta$, $\alpha$ and $\theta$, associated with the center element, i.e., the input is a ($9\times 2 \times S+3$) dimensional vector, where $S$ is the number of cell sites per element. 
    \item The NN is a fully connected network consisting of four hidden layers, each comprising $50$ neurons with ReLU activation functions.
    \item The output layer consists the $6$ components of $\mathbf{\hat{L}}$ (the predicted values of $\mathbf{L}$) and the volume fraction $\hat{v}$ (the predicted value of $v$), i.e., it is a $7$ dimensional vector denoted as $\mathbf{\hat{\Psi}}$ (the predicted value of $\mathbf{\Psi}$). To ensure that the element stiffness matrix remains positive definite, the predicted diagonal components $\{\hat{L}^{(e)}_{00}, \hat{L}^{(e)}_{11}, \hat{L}^{(e)}_{22}\}$ are clipped, from below, to a small positive value (of $10^{-6}$).
\end{enumerate}

The NN is then trained on the $(N_d)$ generated data, i.e., the NN minimizes the mean squared difference between the actual and predicted microstructure properties. The loss function is defined here to be:

\begin{equation}
L_{NN} = \frac{1}{N_d}\big( ||\mathbf{\hat{\Psi}}- \mathbf{\Psi}||^{2}_2 \big)
\label{eq:nn_loss}
\end{equation}

\subsection{Multi-Scale Optimization}
\label{sec:multiscale_optimization}

\begin{figure*}
 	\begin{center}
		\includegraphics[scale=0.7,trim={0 0 0 0},clip]{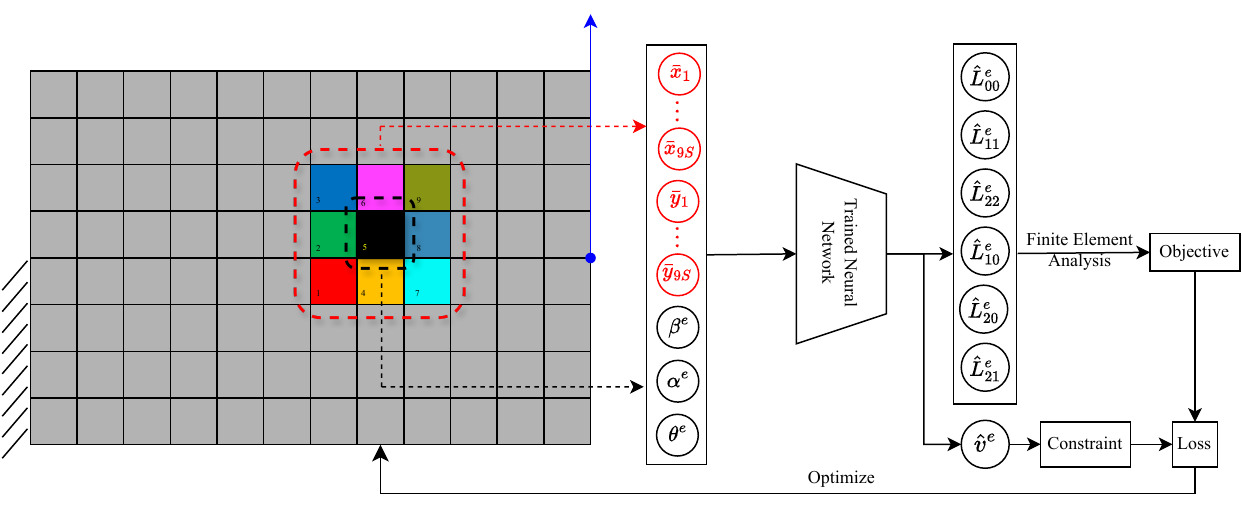}
 		\caption{The multiscale topology optimization framework utilizes the cell site locations of the microstructure and its neighboring microstructures, along with the microstructure's thickness, degree of anisotropy, and orientation to predict the constitutive properties. This prediction is achieved using the trained neural networks, facilitating finite element analysis in multiscale optimization. Note: Each microstructure element comprises $S$ sites, and for a neighborhood of 9 elements, the neural network is provided with an input of $9S$ sites.}
        \label{fig:multi_opt}
	\end{center}
 \end{figure*}

We now consider the proposed multi-scale optimization framework illustrated in \cref{fig:multi_opt}. Recall that each macro element is associated with $S$ cell cites and 3 geometric parameters $\beta^{(e)}$, $\theta^{(e)}$ and $\alpha^{(e)}$. These cell sites and geometric parameters will be optimized to meet a desired objective. To avoid expensive homogenization, the trained NN can be exploited to predict the elasticity response of each element. Specifically, at each step of the optimization process, the current locations of the cell sites of $e$ and its 8 neighbors, together with the current values of 3 geometric parameters of $e$ are extracted. We then apply a smoothing radial filter on the three parameters, i.e., at each element, a weighted average of every parameter about the neighborhood of that element is computed. This filtering process helps promote connectivity by smoothing out abrupt changes across neighboring microstructures.

Using these smoothed parameters, the trained NN is used to estimate  $\mathbf{\hat{L}}^{(e)}$, and therefore $\mathbf{\hat{C}}^{(e)}$.  Given the elasticity matrix $\mathbf{\hat{C}}^{(e)}$ of a macro element, recall that the element stiffness matrix  is defined as \cite{chandrasekhar2023GMTOuNN}: 

\begin{equation}
[K^{(e)}] = \int\limits_{\Omega^{(e)}} [\nabla N^{(e)}]^T[\mathbf{\hat{C}}^{(e)}][\nabla N^{(e)}] d \Omega^{(e)}
\label{eq:K_matrix_FE}
\end{equation}
where $[\nabla N^{(e)}]$ is the gradient of the shape matrix. Once the element stiffness matrices are computed, the global stiffness matrix $\bm{K}$ and force vector $\bm{f}$ are assembled, followed by the solution of the displacement field $\bm{u} = \bm{K}^{-1} \bm{f}$. 
The objective, etc, are then computed followed by an update of all design parameters. These steps are discussed in detail in the remainder of this section.

\subsubsection{Optimization Problem}
\label{sec:Opt_Cons}

We now describe the multi-scale optimization formulation.

\textbf{Design Variables:}  The design variables for each element are represented by $\bm{\zeta}^{(e)} = \{\bm{\Delta}_x^{(e)}, \bm{\Delta}_y^{(e)}, \beta^{(e)}, \alpha^{(e)}, \theta^{(e)}\}$. The complete set of design variables is represented as $\bm{\overline{\zeta}} = \{\bm{\zeta}^{(1)}, \bm{\zeta}^{(2)}, ...., \bm{\zeta}^{(n)} \}$. where $n$ is the number of macro elements.

\textbf{Objective:} 
We consider a simple compliance minimization objective:
\begin{equation}
    \label{eq:min_obj}
         J(\overline {\bm{\zeta}}) = \bm{f}^T \bm{u}
\end{equation}
where $\bm{u}$ the nodal displacements, and  $\bm{f}$ is the imposed load.

\textbf{Volume Constraint:} A  global volume constraint is imposed:
\begin{equation}
    \label{eq:vol_cons}
    g_{V} (\overline {\bm{\zeta}}) \equiv \frac{1}{n} \frac{\sum\limits_{e=1}^{n}\hat{v}^{(e)}} {  v_{max}} - 1 \leq 0
\end{equation}
where $\hat{v}^{(e)}$ is the predicted volume fraction, and $v_{max}$ is the imposed upper bound on the global volume fraction.

Consequently, the optimization problem can be formulated as follows:
\begin{subequations}
	\label{eq:optimization_base_Eqn}
	\begin{align}
		& \underset{\overline {\bm{\zeta}} = \{ \bm{\zeta}_1, \bm{\zeta}_2, \ldots \bm{\zeta}_{n} \}} {\text{minimize}} 
           & J(\overline {\bm{\zeta}}) =  \bm{f}^T \bm{u} \label{eq:optimization_base_objective} \\
		& \text{subject to}
		& \bm{K}(\overline {\bm{\zeta}})\bm{u} = \bm{f}\label{eq:optimization_base_govnEq}\\
		& \text{and} &  g_{V} (\overline {\bm{\zeta}}) \equiv \frac{\sum\limits_{e=1}^{n}\hat{v}^{(e)}} { n v_{max}} - 1 \leq 0  
            \label{eq:optimization_base_volCons} \\
	\end{align}
\end{subequations}

\subsubsection{Loss Function}
\label{sec:method_lossFunction}

The constrained minimization problem \cref{eq:optimization_base_Eqn} is transformed into an unconstrained loss function minimization, using the penalty scheme \cite{nocedal1999numerical}. Specifically, the loss function is defined as:

\begin{equation}
    \mathcal{L}(\bm{\zeta}) = \frac{J(\bm{\zeta})}{J^0} + \gamma g_v(\bm{\zeta})^2 
    \label{eq:lossFunction}
\end{equation}

where $J^0$ is the initial compliance. The constraint penalty parameter $\gamma$  begins with an initial value of $\gamma = 0.1$ and is subsequently incremented by $\Delta \gamma = 0.25$ after each epoch. The gradient-based Adam optimizer \cite{kingma2014adam} is used to minimize \cref{eq:lossFunction}.

\subsubsection{Sensitivity}
\label{sec:method_sensitivity}
A critical aspect of gradient-based optimization is determining the sensitivity, or derivatives, of both the objective function and constraints with respect to the optimization parameters. Here, we harness the automatic differentiation (AD) capabilities of the PyTorch framework \cite{pytorch} to avoid manual sensitivity calculations \cite{chandrasekhar2021auto}. In practical terms, we only need to define the forward expressions, and PyTorch's computational library will compute all necessary derivatives with machine precision \cite{vasilev2019python}.

\section{Numerical Experiments}
\label{sec:results}

In this section, we conduct several experiments to illustrate the method. Without loss of generality, the default parameters are as follows:

\begin{enumerate}
    \item We assume the domain to consist of $40 \times 20$ elements, where each element is assumed to be of unit area.

    \item The material is assumed to have Young's modulus of 1 and Poisson's ratio of 0.3.

    \item The force is assumed to be 1 unit.

    \item Termination criteria include an optimization loss tolerance of $10^{-3}$ or a maximum of 300 iterations.

    \item All experiments were conducted on a MacBook M2 Air with $16$ GB RAM.

\end{enumerate}
Other parameters are provided in the corresponding section.
\subsection{Offline Experiments}
\label{sec:oflline_results}

\subsubsection{Dataset Generation}
\label{sec:data_var}
To establish a mapping between microstructure element parameters and microstructure properties using a surrogate NN, the first step involves acquiring training data. This process entails allowing the $4$ seed points per microstructure element $(\bar{x}, \bar{y})$ to be uniformly distributed within a unit-length element. This distribution is achieved by varying the perturbation from the base grid points $(\Delta_x, \Delta_y)$ within the range of $[-0.225, 0.225]$, ensuring a minimum separation of $0.1$ between neighboring cell sites.

\begin{figure}[H]
 	\begin{center}
		\includegraphics[scale=0.6,trim={0 0 0 0},clip]{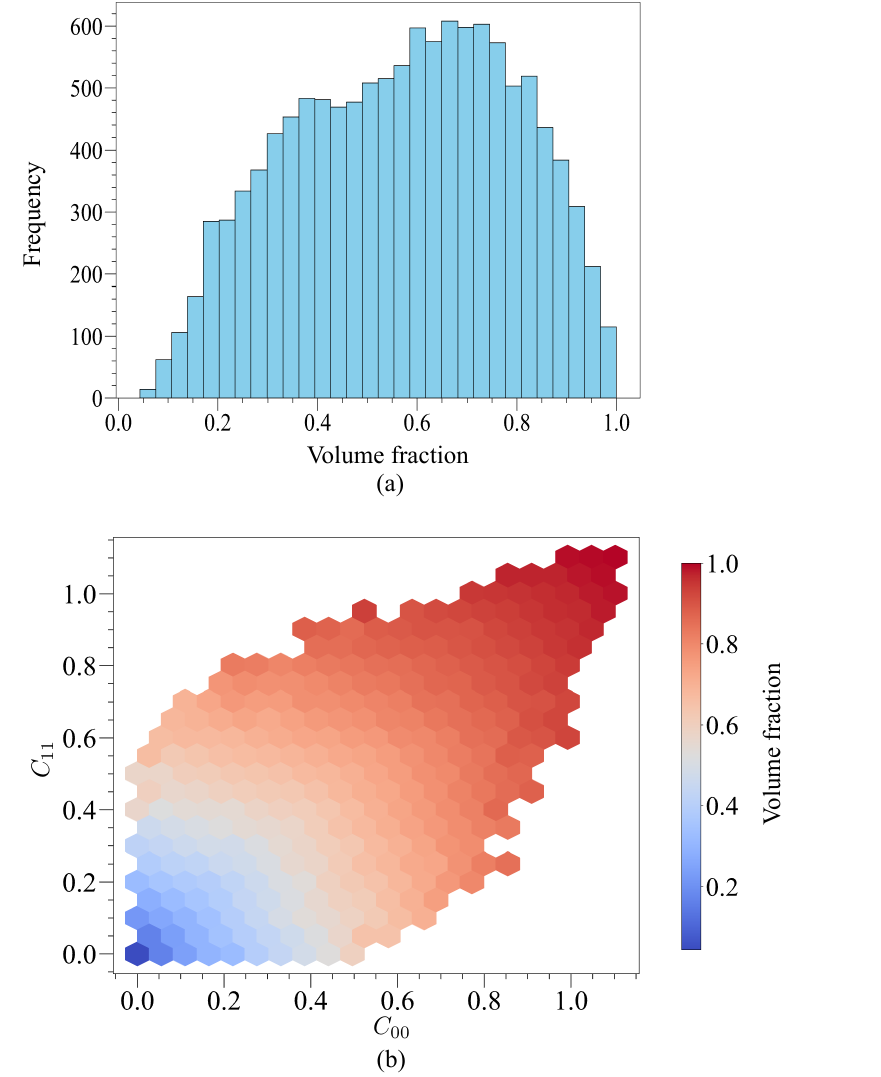}
 		\caption{Distribution of (a) volume fraction and (b) constitutive matrix components $C_{00}$ and $C_{11}$ as a function of microstructure density.}
        \label{fig:data_dist}
	\end{center}
 \end{figure}

Additionally, the parameters thickness ($\beta$), anisotropy ($\alpha$) and orientation ($\theta$) are considered to be in range $[0.3, 3]$, $[1, 3.5]$ and $[0, \pi]$ respectively. These ranges are chosen to encompass a wide variation of volume fractions observed across $12000$ samples of microstructure elements (as depicted in \cref{fig:data_dist} (a)) and to account for anisotropic homogenized elasticity matrix (as shown in \cref{fig:data_dist} (b)). The thickness ($\beta$), anisotropy ($\alpha$) and orientation ($\theta$) were filtered using a filter radius of 3.

 \subsubsection{NN Training }
\label{sec:nn_eval}

\begin{figure}[H]
 	\begin{center}
		\includegraphics[scale=0.3,trim={0 0 0 0},clip]{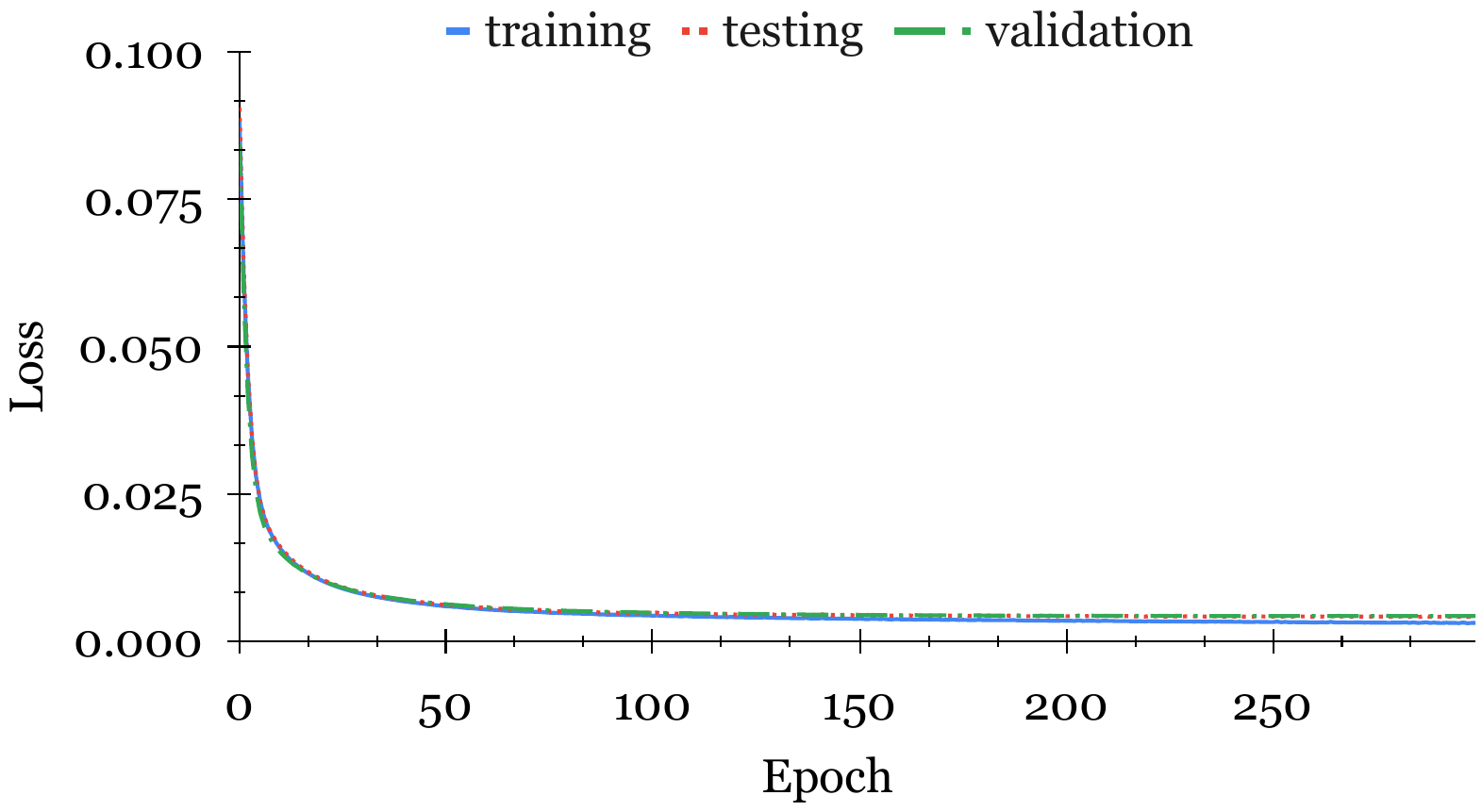}
 		\caption{Convergence of the training, testing, and validation loss.}
        \label{fig:nn_conv}
	\end{center}
 \end{figure}

The training data generated in the previous section serves as input for the surrogate NN, which consists of an architecture outlined in \cref{sec:sur_nn}. The network's output comprises the microstructure volume fraction and Cholesky factors of the homogenized elasticity matrix. The data is split into three sets: training ($N_d = 10000$ data points), validation ($1000$ data points), and testing ($1000$ data points).
During training, the NN aims to minimize the mean squared error between its predicted and actual microstructure properties. This training process is performed for a maximum of $300$ iterations, with a learning rate of $5*10^{-5}$ and a batch size of $64$. The training process is stopped at the maximum iteration number or when the validation loss starts to increase, thus preventing over-fitting. Following this approach, we observe convergence in training, test, and validation loss between actual and predicted volume fraction and homogenized elasticity matrix, as depicted in \cref{fig:nn_conv}. 

\subsection{Multi-Scale Optimization Experiments}
Next, we demonstrate multi-scale optimization through several experiments.

\subsubsection{Validation: Tensile Bar}
\label{sec:expt_validation}
\begin{figure}[H]
 	\begin{center}
		\includegraphics[scale=0.7,trim={0 0 0 0},clip]{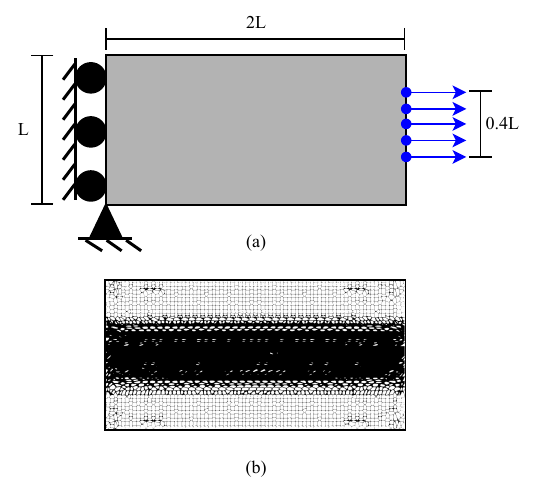}
 		\caption{Validation of the multiscale framework: (a) tensile bar boundary conditions, (b) multiscale design \cite{chandrasekhar2023GMTOuNN} and (c) multiscale porous design. }
        \label{fig:tens_bar}
	\end{center}
 \end{figure}

Consider the tensile bar problem in \cref{fig:tens_bar}(a). The objective is to find the optimal topology that minimizes compliance subjected to a volume fraction $v = 0.4$ as described in \cite{chandrasekhar2023GMTOuNN}.  The compliance of single-scale optimized design reported in reference \cite{chandrasekhar2023GMTOuNN} is $183$. Here, we obtain a multiscale optimized design, illustrated in \cref{fig:val_mid_cant} (c). As one can observe the porous structure resembles an ideal tensile bar, while meeting the minimum porosity imposed (via the parameter $\beta$). Furthermore, near the transition, the porous structures are oriented along the axis, as expected. The compliance of the porous structure is $209$. 

\subsubsection{Comparison}
\label{sec:expt_comp}

\begin{figure}[H]
 	\begin{center}
		\includegraphics[scale=0.7,trim={0 0 0 0},clip]{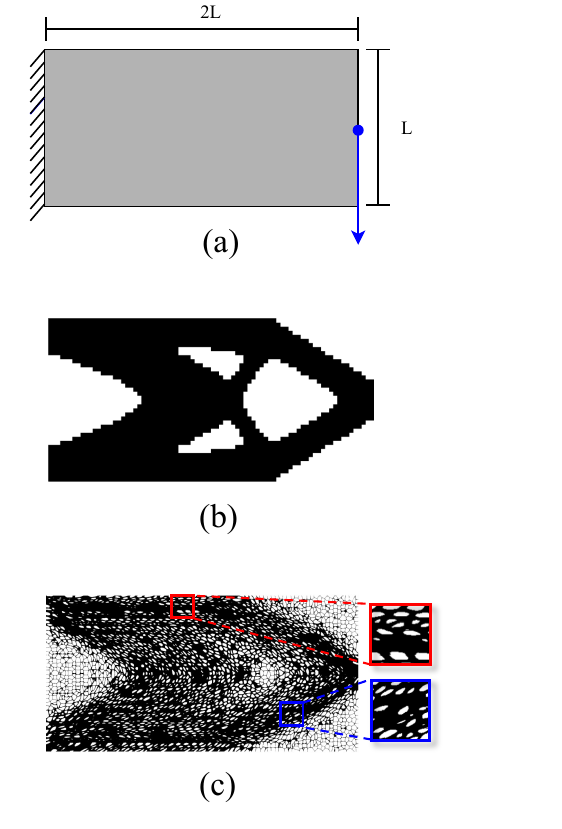}
 		\caption{Validation of the multiscale framework: (a) mid cantilever boundary conditions, (b) single scale design and (c) multiscale porous design. }
        \label{fig:val_mid_cant}
	\end{center}
 \end{figure}

Next consider the mid-cantilever problem in \cref{fig:val_mid_cant}(a). The objective is to find the optimal topology that minimizes compliance subject to a volume fraction $v = 0.6$.  The single-scale optimized design achieved using the code from \cite{andreassen2011efficient} is illustrated in \cref{fig:val_mid_cant}(b); the final compliance is $61.05$. In \cite{wu2017infill}, by imposing a local volume constraint of $v = 0.6$, an optimized porous structure with compliance of $76.86$ was reported. Here, we obtain a multiscale optimized design, illustrated in \cref{fig:val_mid_cant} (c), with a compliance of $68.3$. The computation took $13.7$ seconds. Once again, the computed structure resembles the single-scale structure but deviates from it to meet the porosity constraints.  

To evaluate the accuracy of the NN mapping, we reconstructed the microstructures of the optimized design. We then computed the actual homogenized matrices via FEA of each macro element. Subsequent global analysis yielded a compliance of $74.6$ and a volume fraction of $0.65$. This translates to an error of $8.4\%$ and $6.1\%$ for the compliance and volume fraction, respectively.

\subsubsection{Convergence against Single-Scale}
\label{sec:expt_convergence}

\begin{figure}
 	\begin{center}
		\includegraphics[scale=0.28,trim={0 0 0 0},clip]{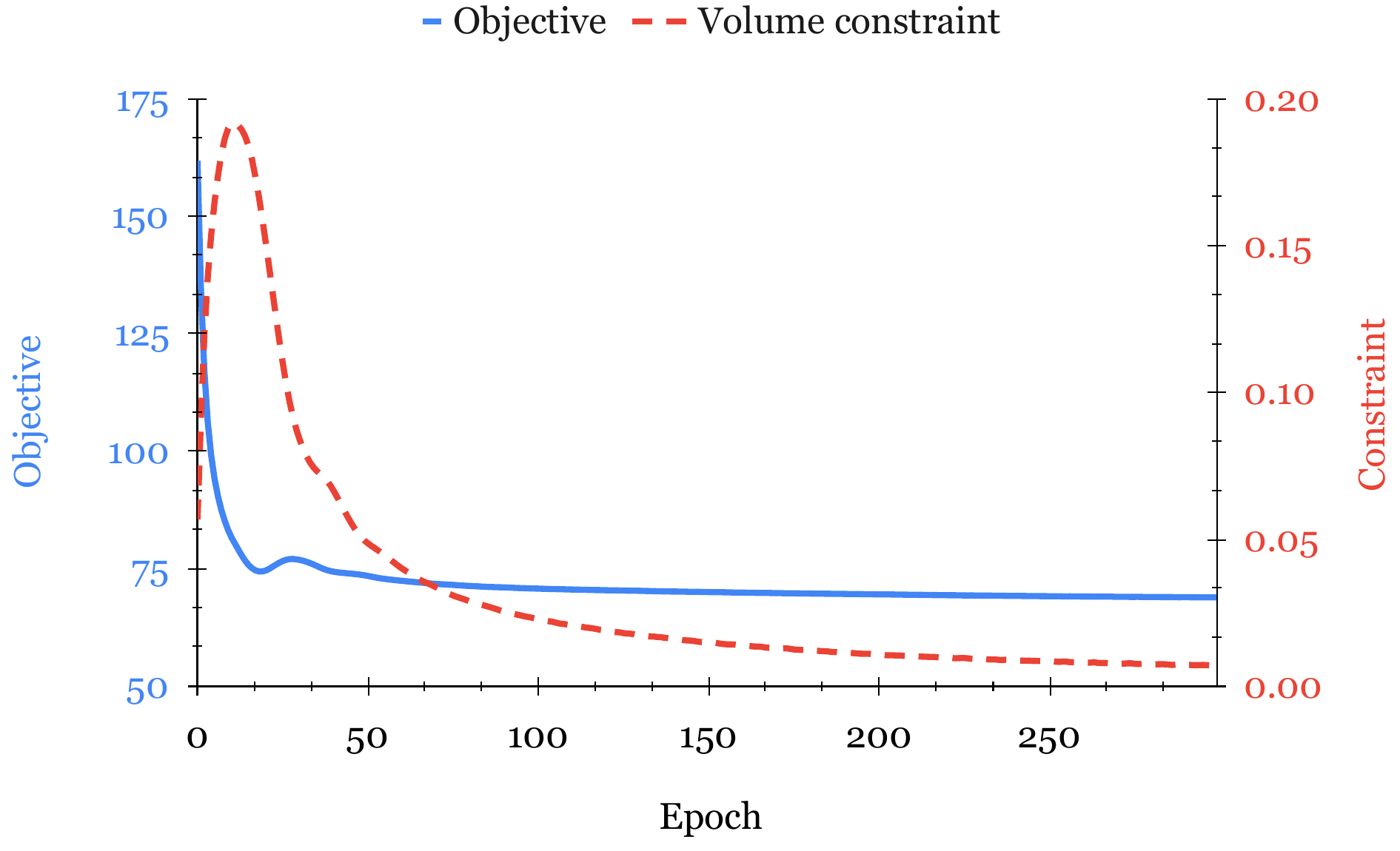}
 		\caption{ Convergence of the objective and constraint during optimization.}
        \label{fig:convergence_graph}
	\end{center}
 \end{figure}

We illustrate the typical convergence of the proposed algorithm for the mid-cantilever problem in \cref{fig:val_mid_cant}(a). The compliance and volume constraint is illustrated in \cref{fig:convergence_graph}. Similar convergence behavior was observed for all other examples. 

\subsubsection{Parameter Variations}
\label{sec:beta_var}
A central hypothesis of our current work is that better multiscale designs can be obtained with a broader range of parameters: thickness, anisotropy, and orientation. To validate these hypotheses, we revisit the problem depicted in \cref{fig:val_mid_cant}(a), enforcing a volume constraint of $0.5$ but considering various restrictions on the parameters.

\textbf{Thickness Parameter: $\beta$}

\begin{figure}[H]
 	\begin{center}
		\includegraphics[scale=0.55,trim={0 0 0 0},clip]{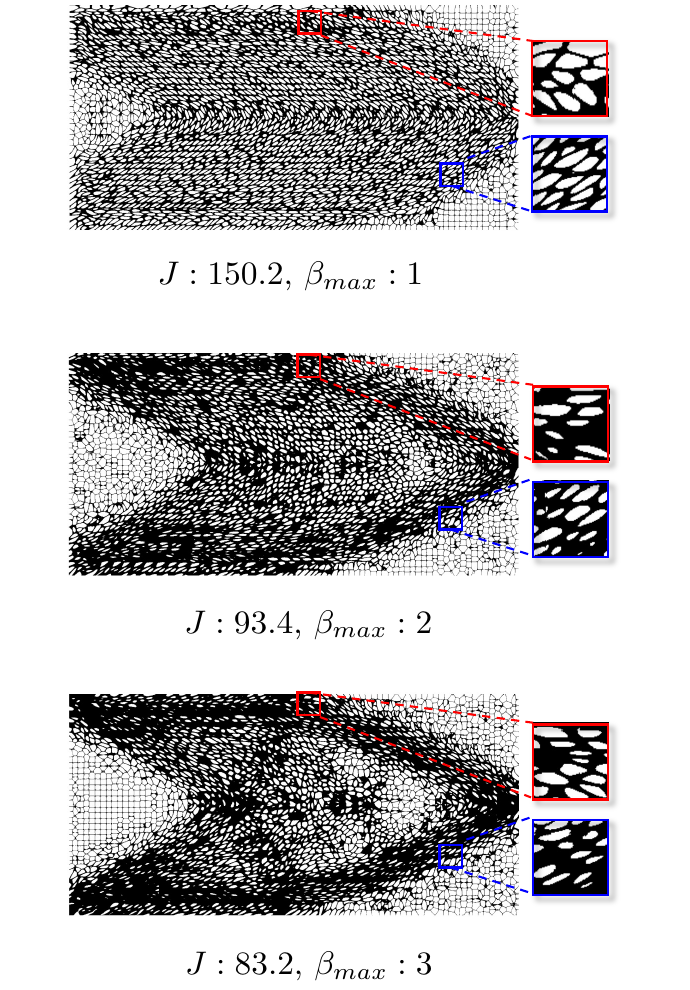}
 		\caption{Impact of thickness parameter on compliance.}
        \label{fig:beta}
	\end{center}
 \end{figure}

In this numerical experiment, we set the lower bound of the thickness parameter at $0.3$ and analyze the impact of thickness on both the objective function and computational time by setting its upper bound to $1.$, $2.$, and $3.$. The resulting topologies, illustrated in \cref{fig:beta}, confirm our expectations: the objective function improves with an increasing range of thickness parameters. Furthermore, the computational time was approximately $13.7 s$, regardless of the range of thickness parameter. 

\textbf{Anisotropy Parameter: $\alpha$}

\begin{figure}
 	\begin{center}
		\includegraphics[scale=0.55,trim={0 0 0 0},clip]{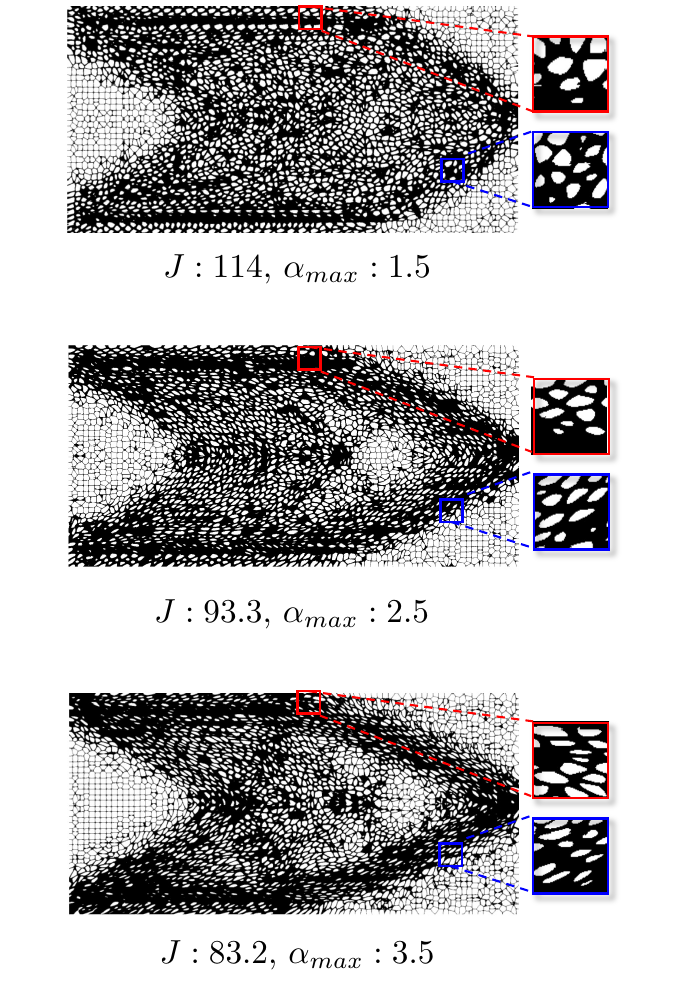}
 		\caption{Varying maximum degree of anisotropy with minimum degree of anisotropy = $1$}
        \label{fig:aniso_var}
	\end{center}
 \end{figure}

In this numerical experiment, we set the lower bound for the degree of anisotropy at $1$ and investigate the influence of anisotropy on both the objective function and computational time by setting its upper bound to $1.5$, $2.5$, and $3.5$. The resulting topologies, as depicted in \cref{fig:aniso_var}, align with our anticipated outcomes: the objective decreases with increasing maximum degrees of anisotropy. Once again, the computational time was unaffected by the change in upper bound. 

 \textbf{Orientation Parameter: $\theta $}

\begin{figure}[H]
 	\begin{center}
		\includegraphics[scale=0.55,trim={0 0 0 0},clip]{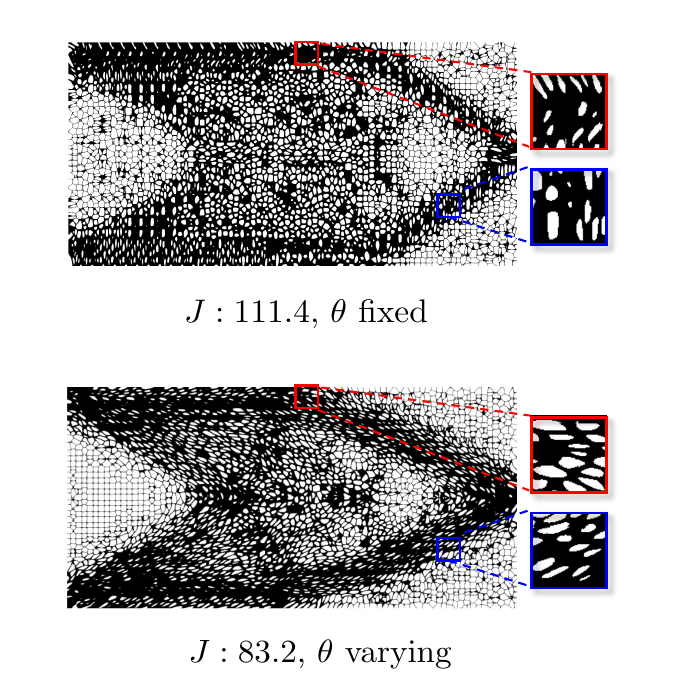}
 		\caption{Impact of orientation on compliance.}
        \label{fig:orient}
	\end{center}
 \end{figure}

Next, we kept the orientation fixed at $\theta =0 $ and optimized the design. As expected, our observations, fixing the orientation results in lower performance as seen in \cref{fig:orient}. 

 \textbf{Volume Fraction Variation}

\begin{figure}[H]
 	\begin{center}
		\includegraphics[scale=0.55,trim={0 0 0 0},clip]{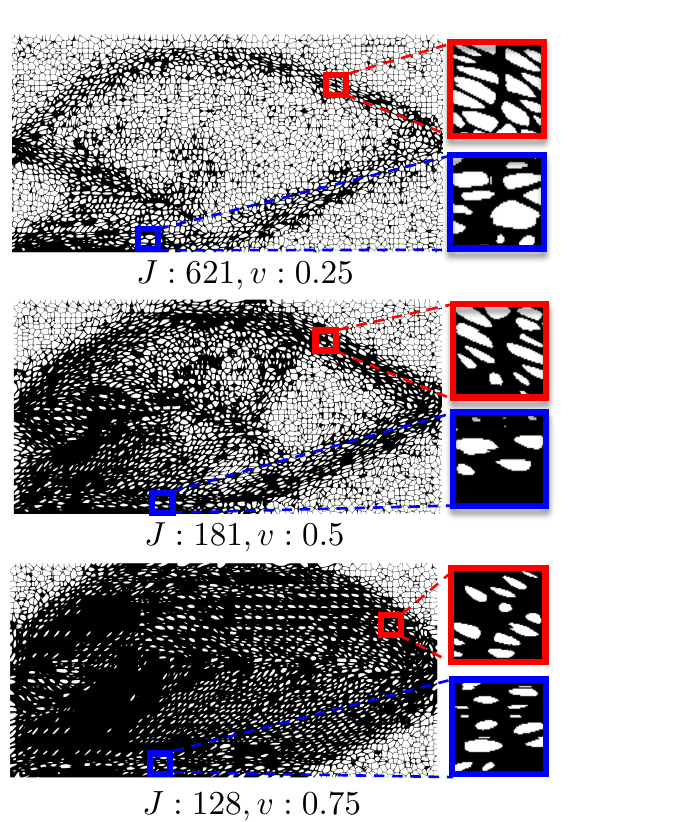}
 		\caption{Trade-off between compliance and volume fraction.}
        \label{fig:pareto}
	\end{center}
 \end{figure}

Finally, we consider the trade-off between the objective (compliance) and constraint (volume fraction) through exploration of the Pareto front is crucial for making informed design choices. In this study, we examined the heel bone problem depicted in \cref{fig:comparison_single_voro} (a). We determined the optimal topologies for various volume fraction constraints. The results are illustrated in \cref{fig:pareto}; as expected, the compliance increases as the volume fraction decreases. To evaluate the accuracy of the NN mapping, we recomputed the actual values using a macroscale FEA/ homogenization approach. The error in compliance and volume fraction was less than $10\%$ in all cases.

\subsection{Is Offline Computation Justified?}
\label{sec:comp_cost}

The offline homogenization and data generation of $12000$ microstructures required $26.4$ minutes, while the training of the surrogate 
NN required $2$ minutes. Thus, the total one-time overhead amounted to $28.4$ minutes. Subsequently, the optimization of the mid-cantilever, for example, (using a grid size of $40 \times 20$) consumed $0.24$ minutes.

Now, let's consider two hypothetical scenarios. The first scenario is a brute-force fine-scale optimization. We note that the grid size of $40 \times 40$ microscale elements is required for each homogenization of the microstructure element. For the single-scale optimization of the mid-cantilever (using a grid size of $40 \times 20$ of microstructure elements), one requires a grid of $40 \times 40 \times 40 \times 20$, which is $1600\times 800 $ elements. One iteration of single-scale optimization with $1600\times 800 $ elements takes $16.62$ minutes. Consequently, the anticipated total optimization time amounts to  $16.62 \times 300$, equating to $4986$ minutes.

Next, consider a scenario where we carry out multiscale optimization but do not rely on offline training, i.e., we carry out concurrent homogenization-based multiscale optimization. Observe that the time required for each homogenization is $26.4/12000$, i.e., $0.132$ seconds. Now, for concurrent homogenization of the mid-cantilever, the task entails homogenization across every element within a $40 \times 20$ grid over $300$ iterations. As a result, the projected optimization duration can be estimated as $0.132 \times 800 \times 300/60$, which equals $528$ minutes.

Thus, the proposed offline NN-based multiscale optimization is computationally far superior. Furthermore, the loss in accuracy in predicted compliance and volume fraction was found to be less than $10 \%$ in all experiments.

\section{Conclusion}
\label{sec:conclusion}

In this paper, we introduced a novel multi-scale topology optimization framework using Voronoi microstructures. Our approach involves an offline homogenization process coupled with NN training to establish a continuous and differentiable mapping of microstructural geometric parameters to constitutive properties. Subsequently, we carried out multiscale optimization to minimize compliance subject to volume constraints.

Our numerical results illustrate that the proposed method offers significant computational advantages over concurrent homogenization, with minimal loss in accuracy. Additionally, we observe that Voronoi microstructures, along with parameters such as thickness anisotropy and orientation, expand the design space. Furthermore, considering neighboring microstructure cell sites during NN training facilitates macroscopic connectivity. Finally, training the NN with Cholesky factors ensures the positive definiteness of the constitutive matrix.

Future research avenues include extending the framework to encompass Voronoi structures found in natural systems, which often serve multiple functions. For example, in bone infills, pores act as conduits for transport while the material provides structural integrity. This could involve a multi-objective formulation integrating compliance and diffusivity to generate bone-like porous structures.
In natural porous materials, the density of Voronoi cell sites varies, whereas in the present formulation, the number of cell sites in a region remains constant. Extending the framework to three dimensions is also a crucial area for future investigation. Additionally, the proposed framework focused on a singular class of microstructures derived from Turing patterns, but the potential for generating a broader spectrum of patterns exists through the application of reaction-diffusion equations \cite{hankins2021methodology}.

\section*{Acknowledgments}
The University of Wisconsin, Madison Graduate School supported this work. 

\section*{Compliance with ethical standards}
The authors declare that they have no conflict of interest.

\section*{Replication of Results}
The Python code is available at \href{https://github.com/UW-ERSL/VoroTO}{github.com/UW-ERSL/VoroTO}

\bibliographystyle{unsrt}  
\bibliography{references}

\end{document}